\documentclass[a4paper, amsfonts, amssymb, amsmath, reprint, showkeys, nofootinbib, twoside]{revtex4-1}
\usepackage[english]{babel}
\usepackage[utf8]{inputenc}

\usepackage[pdftex, pdftitle={Article}, pdfauthor={Author}]{hyperref} % For hyperlinks in the PDF
\usepackage{subfigure}
\usepackage{graphicx}
\usepackage{amsthm}
\usepackage{mathtools}
\usepackage{physics}
\usepackage{xcolor}
\usepackage{graphicx}
\usepackage{adjustbox}
\usepackage{placeins}
\usepackage[T1]{fontenc}
\usepackage{lipsum}
\usepackage{csquotes}
\usepackage{amssymb}
\usepackage{bbm}
\usepackage[left=16mm,right=16mm,top=35mm,columnsep=15pt]{geometry}
\begin{document}
\title{Local assortativity affects the synchronizability of scale-free network}

\author{Mengbang Zou}
\author{Weisi Guo}
\email[Correspondence email address: ]{weisi.guo@cranfield.ac.uk}% Your name
\affiliation{Department of Aerospace Engineering, Cranfield University, Cranfield, MK43 0AL, United Kingdom}

\date{\today} % Leave empty to omit a date

\begin{abstract}
Synchronization is critical for system level behaviour in physical, chemical, biological and social systems. Empirical evidence has shown that the network topology strongly impacts the synchronizablity of the system, and the analysis of their relationship remains an open challenge. We know that the eigenvalue distribution determines a network's synchronizability, but analytical expressions that connect network topology and all relevant eigenvalues (e.g., the extreme values) remain elusive.  

Here, we accurately determine its synchronizability by proposing an analytical method to estimate the extreme eigenvalues using perturbation theory. Our analytical method exposes the role global and local topology combine to influence synchronizability. We show that the smallest non-zero eigenvalue $\lambda^{(2)}$ which determines synchronizability is estimated by the smallest degree augmented by the inverse degree difference in the least connected nodes $\lambda^{(2)} \simeq \min{\bigg((k_i-1) + \sum_{j \ne i}\frac{A_{ij}^2}{(k_i-1)-k_{j}} \bigg|_{k_i=k_{\rm min}}\bigg)}$. From this, we can conclude that there exists a clear negative relationship between $\lambda^{(2)}$ and the local assortativity of nodes with smallest degree values. We validate the accuracy of our framework within the setting of a Scale-free (SF) network and can be driven by commonly used ODEs (e.g., 3-dimensional Rosler or Lorenz dynamics). From the results, we demonstrate that the synchronizability of the network can be tuned by rewiring the connections of these particular nodes while maintaining the general degree profile of the network. 
\end{abstract}

\keywords{perturbation theory, synchronizability, complex network, local assortativity}

\maketitle

\section{Introduction} \label{sec:introduction}
Synchronization, as a collective phenomenon of dynamically coupling units, generally exists in different fields such as power grids \cite{dorfler2013synchronization}, wirelss communication networks  \cite{9169833}, neural networks \cite{chen2020new}, etc. Realizing that network topology of system plays an important role in system's behavoirs, the relationship between network topology and synchronizability has attracted a lot of attention in recent years \cite{zhao2006relations, hong2004factors, duan2007complex, aguirre2014synchronization, rungta2017network}. According to intuitive experience, some network topology characteristics are proposed as indicators for synchronizability such as the betweenness centrality \cite{hong2004factors}, degree correlation \cite{di2007effects}, etc. One problem needs to be pointed out is that when analysing the relationship between synchronizability and topology characterictics, some parameters like number of nodes $N$ of the network, rewiring probability $p$ of SW (small-world) networks need to be adjusted, which would cause other network topology characteristics changing, such as average distance, clustering coefficients, etc. The direct relationship between synchronizability and a given topology characterisctic is not clear when other network topology characteristics keep varying. Besides, some topology characteristics provide indicators of synchronization in a network class but fail in other network classes \cite{rungta2017network}. The previous simulation experiments could reveal the relationship between synchronizability and network topology characteristics in some situation, but are far from clearly explaining it or mathematically abstracting it. It is important for researchers to uncover the behavior of empirical phenomena through experiments and data analysis. More importantly, we need to develop theories that abstract such behavior mathematically and explain the mechanism behind these phenomena \cite{popper2005logic}. The master stability function relates the global synchronizability to the spectral properties of the Laplacian matrix of the network, which provides the objective criterion for synchronizability \cite{pecora1998master}. For determined self-dynamics function and coupling dynamics function, the global synchronizability of the network is determined by the spectral properties of the network. Based on this analysis framwork, the analysis of synchronizability of complex network could be coverted to analyze the bounds of the extreme eigenvalues \cite{arenas2008synchronization} \cite{sarkar2018spectral}. There are some main results in previous works in different network models, here we mainly discuss SF network.

\begin{figure*}[htbp]
\centering
\resizebox*{15cm}{!}{\includegraphics{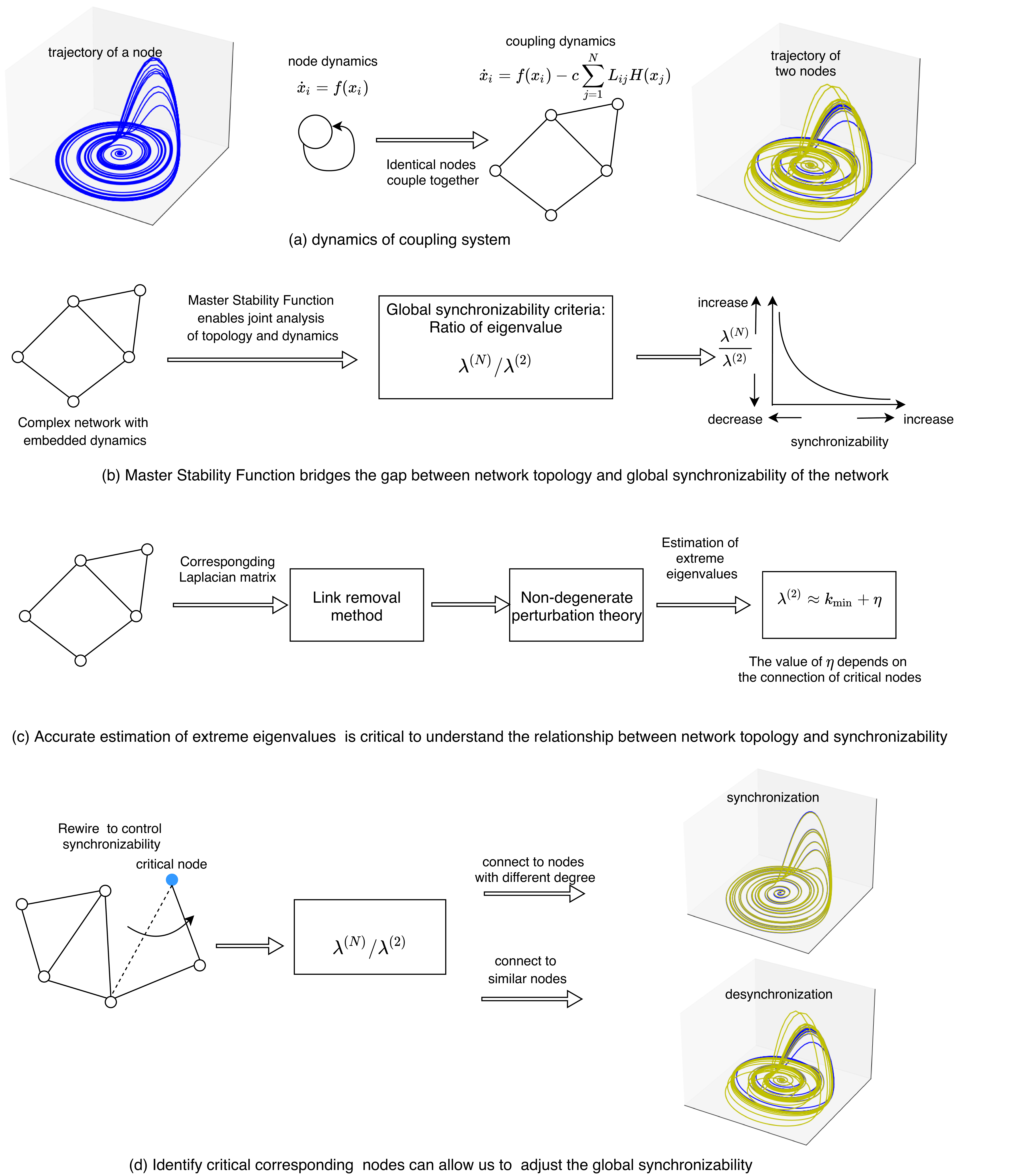}}
\caption{This figure shows our proposed analysis framework of global synchronizability. (a) describes the dynamics of a node and the dynamics of coupled nodes. (b) shows that Master Stability Function provides us a criteria of global synchronizability, the ratio of extreme eigenvalues. (c) uses non-degenerate perturbation theory to estimate extreme eigenvalues. (d) proposes a strategy to control the global synchronizability by rewiring connections of critical nodes.}
\label{paperstructure}
\end{figure*}

In a SF network with large minimum degree or random enough network, the extreme eigenvalues can be bounded by the mean degree, minimum degree and maximum degree \cite{zhou2006universality}.  Perturbation analysis of spectral of SF network shows that the maximum eigenvalue is approximately equal to the maximum degree \cite{kim2007ensemble}, but the smallest non-zero eigenvalue $\lambda^{(2)}$ is ensemble averageable by themselves. The difficulty of approximating $\lambda^{(2)}$ by perturbation theory is the degeneration of eigenvectors which is caused by the existing of a large number of nodes with the smallest degree. To solve this problem, degenerate perturbation theory is employed to estimate eigenvalues and location of eigenvectors of random network \cite{hata2017localization}. Though the degenrate perturbation theory can estimate eigenvalues accurately, it requires the gobal information of the network and introduction of new eigenvectors for several times to solve the degeneration at the first order perturbation \cite{hata2017localization}. This will increase the computation complexity in estimation. Besides, it is difficult for us to get any knowledge about the relationship between extreme eigenvalues and network topology characteristics through the complicated process of calculation. Degenerate perturbation theory is an effective method to estimate eigenvalues, but it is difficult to get an analytic equation to help us understand the synchronizability of the network by this method. Therefore, how to avoid the degeneration of eigenvectors and get a clear analytic equation when estimating the smallest non-zero eigenvalue is necessary. To solve this problem, a link removal method is used to avoid the degeneration of eigenvectors. We prove that the link removal of the node with smallest degree has little effect on the smallest non-zero eigenvalue of the network. The new network after link removal only have one node with smallest degree, which means that the non-degenerate perturbation theory can used to estimate the smallest non-zero eigenvalue of the new network. The non-degenerate perturbation theory only requires the local information of smallest nodes (node of the minimum degree) and the calculation complexity is much lower comparing with the degenerate perturbation theory.

The contribution of this paper is that we propose an analytic framework to get a clear analytic equation when estimating the smallest non-zero eigenvalue of the network (The analytic framework is shown in Fig.(\ref{paperstructure})). The analytic equation points out how the network topology affects the synchronizability of the network. Firstly, the analytic equation shows that the smallest non-zero eigenvalue $\lambda^{(2)}$ is mainly determined by the minimum degree of the network and the connection of smallest nodes in a SF network. Second, according to the analytic equation, there exists relationship between $\lambda^{(2)}$ and local assortativity of these nodes. Besides, the analytic equation instructs us how to strengthen or weaken the synchronizability of the network by reconnecting links among nodes with prescribed degree profile. The assumption of this paper is that the network used in this paper is SF network with large minimum degree and the coupling nodes in the network have identical dynamics.

This paper is organized as follows. In Sec. \ref{sec:synchronization}, we introduce the general dynamics model for networked system and the criteria to characterize synchronicability provided by master stability function method. In Sec. \ref{sec:perturbation}, we propose an analytical method to estimate extreme eigenvalues in SF network by perturbation theory and analyze the relationship between $\lambda^{(2)}$ and local assortativity. In Sec. \ref{sec:results}, Rossler system on different SF network is used to verify our theory. At last, we make a conclustion and discuss about the future direction.

\begin{table}[b]
\caption{\label{tab:table1}
List of symbols used in this paper}
\begin{ruledtabular}
\begin{tabular}{cccccccc}
\textbf{Symbol}&\textbf{Describtion}\\
\hline
$x_i$& state of node $i$\\
$f(\cdot)$  &self-dynamics function\\
$H(\cdot)$  &inner coupling function \\
$c$&coupling strength  \\
$\textbf{L}$& Laplacian matrix of the network \\
$\textbf{D}$& diagonal matrix of degrees \\
$\textbf{A}$& adjacent matrix \\
$\zeta_i$&the variation on node $i$\\
$\gamma$ & $\gamma=c*\lambda^{(i)}$\\
$S$ & the synchronized region\\
$\Omega_{\rm max}$&the largest Lyapunov exponent\\
$L_{ij}$& element of matrix $\textbf{L}$ \\
$A_{ij}$& element of matrix $\textbf{A}$ \\
$\lambda^{(i)}$ & eigenvalue of matrix $\textbf{L}$  \\
$k_{i}$& degree of node $i$ \\
$Df(s)$ & the Jacobian matrix of function $f$ at $s$ \\
$DH(s)$ & the Jacobian matrix of function $H$ at $s$ \\
$\epsilon$& the expansion parameter which tends to be small \\
$\vec{\xi}^{(i)}$ & eigenvector corresponding to $\lambda^{(i)}$\\
$\xi_{\alpha}^{(i)}$ & the $\alpha$-th element of eigenvector $\vec{\xi}^{(i)}$\\
$\vec{\xi}^{(i)tr}$ & the transpose of $\vec{\xi}^{(i)}$\\
$\delta_{ij}$&Kronecker delta function\\
$\rho_i$&local assortativity of node $i$\\
$r$&assortativity of the network\\
$\theta_i$&degree difference between node $i$ and its neighbours
\end{tabular}
\end{ruledtabular}
\end{table}  %I believe leaving the sections in separate files is more organized, change it if you desire 
\section{Network Synchronization} \label{sec:synchronization}
Consider a dynamical complex network with $N$ coupled identical nodes (symbols used in this paper are shown in Table (\ref{tab:table1})), described by
\begin{equation}\label{equ1}
    \dot{x_i}=f(x_i)-c\sum^N_{j=1}L_{ij}H(x_j),  i =1, 2,3,...N,
\end{equation}
where $x_i=(x_i^{(1)}, x_i^{(2)}, ..., x_i^{(n)}) \in \mathbb{R}^n$ is the state vector of node $i$, $f(\cdot):\mathbb{R}^n\xrightarrow[]{}\mathbb{R}^n$ controls self dynamics of node $i$, $c>0$ is the coupling strength, $H(\cdot):\mathbb{R}^n\xrightarrow{}\mathbb{R}^n$ is the inner coupling function.  $\rm \textbf{L}$ is the Laplacian matrix of the network. $\rm \textbf{L}=\rm \textbf{D}-\rm \textbf{A}$, where $\rm \textbf{A}$ is the adjacent matrix and $\rm \textbf{D} = {\rm diag}\{k_1,\cdots, k_N\}$ is the diagonal matrix of degrees. If an edge exists between node $i$ and node $j$, $A_{ij}=A_{ji}=-1$. Otherwise, $A_{ij}=0$. The matrix $\rm \textbf{L}$ satisfies $L_{ii}=-\sum^N_{j=1, j\ne i}L_{ij}, i=1,2,..., N.$ If the graph is connected, then $\rm \textbf{L}$ is irreducible. Zero is an eigenvalue of $\rm \textbf{L}$ with multiplicity 1 and all other eigenvalues are strictly positive, denoted by
\begin{equation}
    0 = \lambda^{(1)}<\lambda^{(2)}\le \lambda^{(3)} \le ... \le \lambda^{(N)}.
\end{equation}
The nodes are labeled in increasing order of their degrees $k_i$, such that $k_{\rm min}=k_1 \le k_2 \le \cdots \le k_N = k_{\rm max}$. The average degree of the network is $<k>=\sum_{i=1}^Nk_i/N$.
The stability of the synchronized manifold $x_1=x_2=...=x_N$ can be determined by the master stability function \cite{pecora1998master}

\begin{equation}\label{equ3}
    \dot{\zeta} = [Df(s)+\gamma DH(s)]\zeta,
\end{equation}

where $\zeta$ is the collection of variations and $\zeta=(\zeta_1, \zeta_2,...,\zeta_N)$. $\zeta_i$ is the variation on the $i$-th node. $Df(s)$ and $DH(s)$ are the Jacobian matrix of functions $f$ and $H$ at $s$. The largest Lyapunov exponent $\Omega_{\rm max}$ of the system (\ref{equ1}) could be calculated from (\ref{equ3}). $\Omega_{\rm max}$ is a function of $\gamma$, where $\gamma=c\lambda^{(i)}$. The region $S$ of $\gamma$ which makes $\Omega_{\rm max}$ negative is called the synchronized region. If the eigenvalue $\lambda^{(i)}$ of matrix $A$ satisfies 
\begin{equation}
    c\lambda^{(i)} \in S, i =2,3, \cdots N,
\end{equation}
then (\ref{equ1}) is asymptotically stable. The synchronized region $S$ could be an unbounded region $(-\infty, \gamma_1)$, a bounded region $(\gamma_1, \gamma_2)$, an empty set and a union of several subregions. In this paper, we mainly consider the the bounded region.

When the synchronized region $S=(\gamma_1, \gamma_2)$, where $\gamma_1, \gamma_2$ are both negative real numbers, the eigenvalues of matrix $\rm \textbf{L}$ need to satisfy
\begin{equation}\label{equ5}
    c\lambda^{(N)}>\gamma_1, c\lambda^{(2)}<\gamma_2
\end{equation}
to make the synchronized region asymptotically stable. (\ref{equ5}) can be written as 
\begin{equation}
\frac{\lambda^{(N)}}{\lambda^{(2)}}<\frac{\gamma_1}{\gamma_2}.
\end{equation}
The ratio of $\frac{\lambda^{(N)}}{\lambda^{(2)}}$ characterizes the synchronizability in this case.

\section{Analyse synchronizability by purterbation theory}
\label{sec:perturbation}
Since the synchronizability of networks depend on the extreme eigenvalues $\lambda^{(2)}$ and $\lambda^{(N)}$, we use nondegenerate perturbation theory to estimate $\lambda^{(N)}$. Similar perturbation methods are used in \cite{kim2007ensemble} \cite{hata2017localization} to estimate eigenvalues and eigenvectors\footnote{except the eigenvalue $\lambda^{(1)}=0$ and its corresponding eigenvector, which have exceptional characteristics and is excluded from the analysis \cite{hata2017localization}}. 

First we introduce the expansion parameter $\epsilon=<k>^{-1}$. The Laplacian matrix $ \textbf{L}$ could be rewritten as $\rm \textbf{L}= \textbf{L}_0+\epsilon  \textbf{L}_1$, where $ \textbf{L}_0=\textbf{D}$ and $ \textbf{L}_1=-<k>\textbf{A}$. $\vec{\xi}^{(i)} = \{\xi_1^{(i)}, \xi_2^{(i)} \cdots \xi_N^{(i)}\}^{T}$ represents the Laplacian eigenvetor of the $i$ th mode and $\lambda^{(j)}$ is the corresponding eigenvalue. Since $\rm \textbf{L}$ is a real symmetric matrix, the eigenvectors can be orthonormalized as $\sum_{\alpha=1}^N \xi_{\alpha}^{(i)}\xi_{\alpha}^{(j)}=\delta_{i,j}$, where $\delta_{i, j}$ is the Kronecker delta function.  $\rm \textbf{L}_0$ is considered as an unperturbed matrix and $\rm \textbf{L}_1$ is considered as a perturbation. We can get 
\begin{equation}\label{equ7}
    \sum_{\beta=1}^NL_{\alpha\beta}\xi_{\beta}^{(i)}=
    \lambda^{(i)}\xi_{\alpha}^{(i)}.
\end{equation}
We expand the eigenvector and eigenvalues with $\epsilon$ as
\begin{equation}\label{equ8}
\begin{split}
    &\vec{\xi}^{(i)} = \vec{\xi}^{(i)}_0 + \epsilon\vec{\xi}^{(i)}_1+\epsilon^2\vec{\xi}^{(i)}_2+\cdots, \\
    &\lambda^{(i)} = \lambda^{(i)}_0 + \epsilon \lambda^{(i)}_1 + \epsilon^2\lambda^{(i)}_2+\cdots.
\end{split}
\end{equation}
We assume that the unperturbed eigenvectors are orthonormalized and the higher-order perturbation vectors are orthogonal to the unperturbed eigenvectors. According to (\ref{equ7}) and (\ref{equ8}), we can obtain the following equations up to $O(\epsilon^2)$\footnote{the value of $\epsilon$ tends to be small and second order approximation is sufficient}:
\begin{equation}\label{equ9}
    \begin{split}
        &(\textbf{L}_0-\lambda_0^{(i)})\vec{\xi}^{(i)}_0=0,\\
        &(\textbf{L}_0-\lambda_0^{(i)})\vec{\xi}^{(i)}_1=(\lambda^{(i)}_1- \textbf{L}_1)\vec{\xi}^{(i)}_0,\\
        &(\textbf{L}_0-\lambda_0^{(i)})\vec{\xi}^{(i)}_2=(\lambda^{(i)}_1- \textbf{L}_1)\vec{\xi}^{(i)}_1 + \lambda^{(i)}_2\vec{\xi}^{(i)}_0,\\
        &(\textbf{L}_0-\lambda_0^{(i)})\vec{\xi}^{(i)}_3=(\lambda^{(i)}_1- \textbf{L}_1)\vec{\xi}^{(i)}_2 + \lambda^{(i)}_2\vec{\xi}^{(i)}_1+\lambda^{(i)}_3\vec{\xi}^{(i)}_0.
    \end{split}
\end{equation}
It is easy to get 
\begin{equation}\label{equ10}
    \begin{split}
        &\lambda^{(i)}_0=k_{i},\\
        &\lambda^{(i)}_1=-<k>A_{i i}=0.
    \end{split}
\end{equation}
Furthermore, it can be deduced 
\begin{equation}\label{equ11}
    \begin{split}
        &\lambda^{(i)}_2=\vec{\xi}_0^{(i)tr}\textbf{L}_1\vec{\xi}^{(i)}_1,\\
        &\lambda^{(i)}_3=\vec{\xi}_0^{(i)tr}\textbf{L}_1\vec{\xi}^{(i)}_2,
    \end{split}
\end{equation}
where $\vec{\xi}_0^{(i)tr}$ is the transpose matrix of $\vec{\xi}_0^{(i)}$. The first-order and second-order correction $\vec{\xi}^{(i)}_1, \vec{\xi}^{(i)}_2$ of the $i-\rm th$ vector can be obtained by 
\begin{equation}\label{equ12}
    \begin{split}
        &\vec{\xi}^{(i)}_1=\sum_{j \ne i}\frac{\vec{\xi}^{(j)tr}_0\textbf{L}_1\vec{\xi}^{(i)}_0}{\lambda^{(i)}_0-\lambda^{(j)}_0}\vec{\xi}^{(j)}_0,\\
        &\vec{\xi}^{(i)}_2=\sum_{h \ne i}\frac{\vec{\xi}^{(h)tr}_0\textbf{L}_1\vec{\xi}^{(i)}_1}{\lambda^{(i)}_0-\lambda^{(h)}_0}\vec{\xi}^{(h)}_0.
    \end{split}
\end{equation}
(\ref{equ12}) reveals that $\vec{\xi}^{(i)}_1$ is determined by the nodes connected to node $i$ which are called the first-order neighbour nodes of node $i$. $\vec{\xi}^{(i)}_2$ is determined by the nodes connected to neighbour nodes of node $i$. These nodes are called the second-order neighbour nodes of node $i$. Furthermore, we can get 
\begin{equation}
    \lambda^{(i)}_2=\sum_{i \ne j}\frac{A_{ij}^2}{k_{i}-k_{j}}.
\end{equation}
That is to say $\lambda^{(i)}_2$ is determined by the first-order neighbour nodes of node $i$. Also, (\ref{equ11}) reveals that $\lambda^{(i)}_3$ is determined by the second-order neighbour nodes of node $i$. $\lambda^{(i)}_3=0$, since the second-order neighbour nodes of node $i$ are not directly connected with node $i$. In a similar way, we can deduce that $\lambda^{(i)}_3, \lambda^{(i)}_4...=0$. Therefore, according to our perturbation analysis, $\lambda^{(i)}$ is mainly determined by its first-order neighbour nodes and almost not affected by other nodes. 

So $\lambda^{(N)}$ could be estimated by perturbation expansion to second order as
\begin{equation}
    \lambda^{(N)} \simeq k_N + \sum_{i \ne j}\frac{A_{Nj}^2}{k_N-k_{j}}.
\end{equation}

The second-order term can be expanded as $\sum_j(A_{Nj})^2(\frac{1}{k_N}+\frac{k_j}{k_N^2}+\cdots)=1+k_N^{\rm ave} / k_N + \cdots$, where $k_N^{\rm ave}$ is the average degree of the nearest neighbours of node $N$. For large $N$ $k_N^{\rm ave}/k_N \ll 1$ \cite{kim2007ensemble}. Therefore, $\lambda^{(N)} \simeq k_N + 1$, which means that $\lambda^{(N)}$ is mainly decided by $k_{\rm max}$. On the other hand, $\lambda^{(2)}$ should be estimated by degenerate theory \cite{hirschfelder1974degenerate}, since several nodes may have the same node degree with $\lambda^{(2)}$ in SF network. However, the difficulty is that some eigenvectors remain degenerate after several estimation steps, so the eigenvectors and eigenvalues can not be determined from (\ref{equ9}). To solve this problem, we remove one link from the node with the lowest degree to get a new network in which $k_{\rm min}-1=k_1 < k_2 \le \cdots \le k_N = k_{\rm max}$. So the non-degenerate perturbation theory could be used to estimate $\lambda^{(2)}$ of the new network. If the effect of link removal on eigenvalues is very small, then we think that $\lambda^{(2)}$ of the new network extremely approaches to the original one. Therefore, we can obtain $\lambda^{(2)}$ of the original network by estimating $\lambda^{(2)}$ of the new network.

\begin{figure}[htbp]\label{link_removal}
\centering
\resizebox*{8cm}{!}{\includegraphics{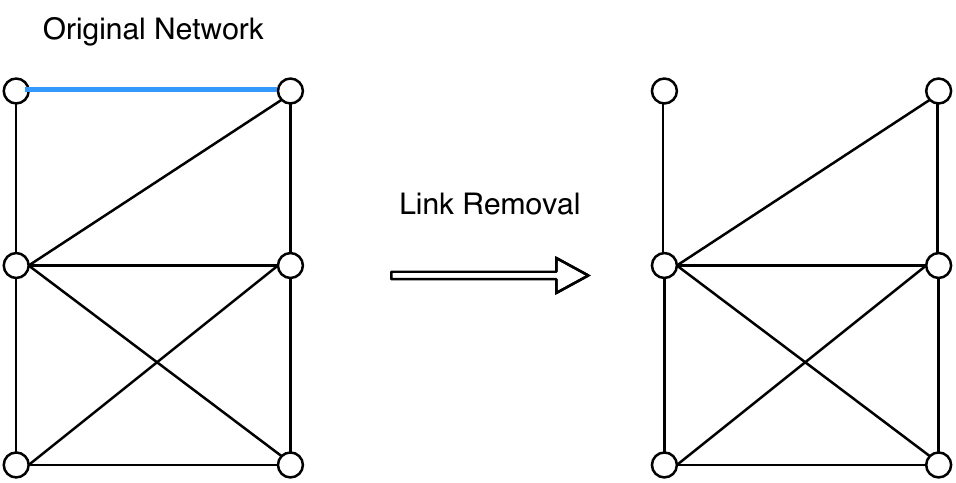}}
\caption{The link removal method is removing the edge between the node with the smallest degree and its neighbour. We can get a network by applying the link removal method to the original network. The smallest non-zero eigenvalues of the original network and the new network is almost the same.}
\label{linkremovalmethod}
\end{figure}

The link removal method is shown in Fig. (\ref{linkremovalmethod}). Here we use perturbation theory \cite{He2019} to analyze the effect of link removal on Laplacian eigenvalues. The change of eigenvalues can be estimated by

\begin{equation}\label{equ13}
    ( \textbf{L}+\Delta  \textbf{L})(\vec{\xi}^{(i)}+\Delta \vec{\xi}^{(i)}) = (\lambda^{(i)}+\Delta\lambda^{(i)})(\vec{\xi}^{(i)}+\Delta \vec{\xi}^{(i)}),
\end{equation}
where $\Delta  \textbf{L}, \Delta \vec{\xi}^{(i)}, \Delta\lambda^{(i)}$ represent the changes in $L,\vec{\xi}^{(i)}, \lambda^{(i)}$. Multiplying (\ref{equ13}) by the transpose of $\vec{\xi}^{(i)}$, $\vec{\xi}^{(i)tr}$, we can get
\begin{equation}
    \Delta\lambda^{(i)}=\frac{\vec{\xi}^{(i)tr}\Delta  \textbf{L}\vec{\xi}^{(i)}+\vec{\xi}^{(i)tr}\Delta  \textbf{L}\Delta \vec{\xi}^{(i)}}{\vec{\xi}^{(i)tr}\vec{\xi}^{(i)}+\vec{\xi}^{(i)tr}\Delta \vec{\xi}^{(i)}}.
\end{equation}

For a large complex network, it is reasonable to assume that the removal of only a link has small effects on the network as well as the eigenvector $\vec{\xi}^{(i)tr}$, which means that $\Delta \vec{\xi}^{(i)tr} \simeq 0$ \cite{PhysRevE.81.046112} \cite{PhysRevLett.97.094102}.
If the link between node $k$ and node $m$ is removed, then 
\begin{equation}
    \Delta \lambda^{(i)}\simeq \frac{2\xi^{(i)}_k\xi^{(i)}_m-\xi^{(i)}_k\xi^{(i)}_k-\xi^{(i)}_m\xi^{(i)}_m}{\vec{\xi}^{(i)tr}\vec{\xi}^{(i)}}.
\end{equation}

\begin{figure*}[htbp]
\centering
\resizebox*{14cm}{!}{\includegraphics{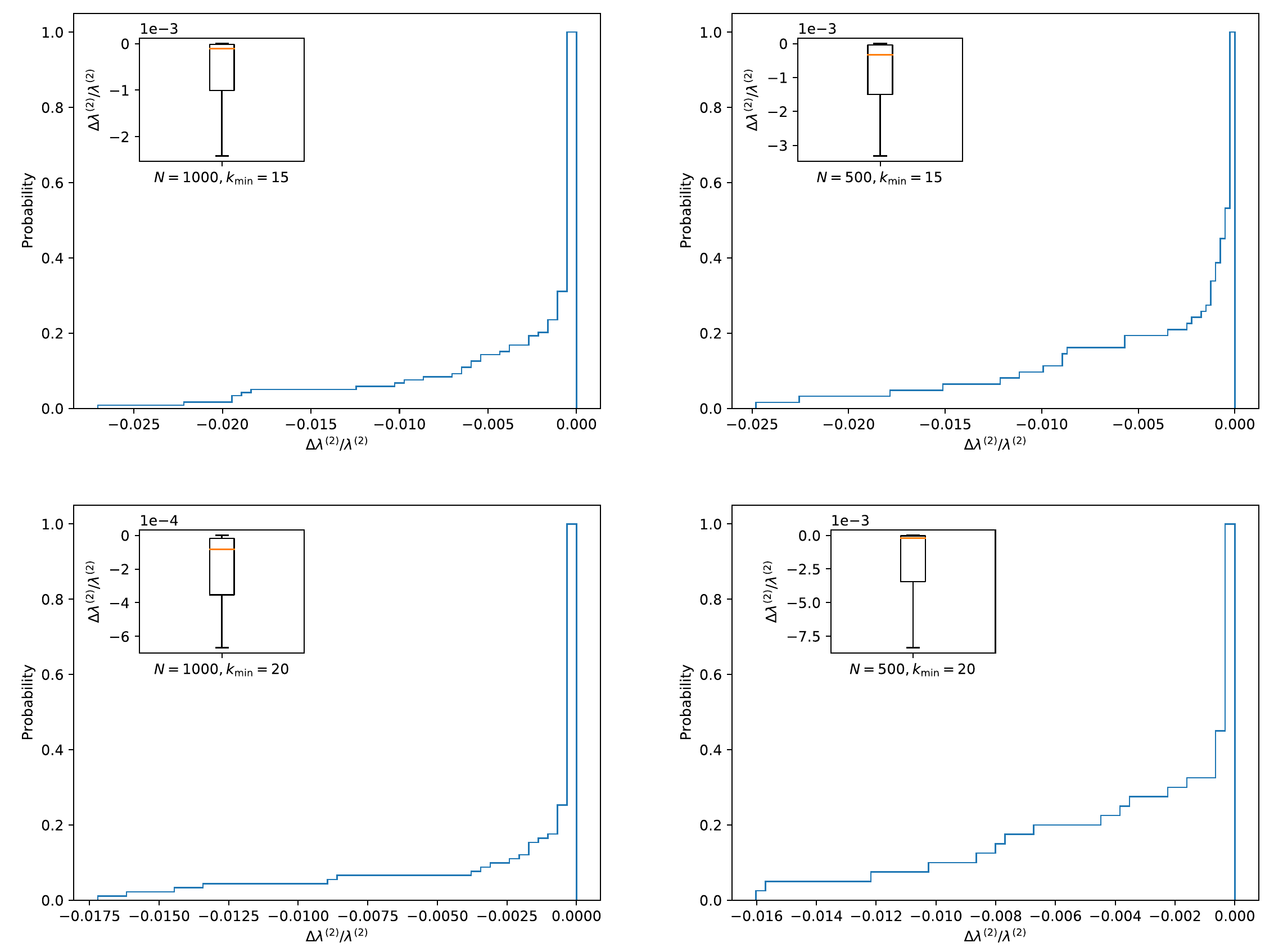}}
\caption{These four figures demonstrate the cumulative distribution of $\Delta \lambda^{(2)}/\lambda^{(2)}$ of different networks. The box-plots show the extrem values and median of $\Delta \lambda^{(2)}/\lambda^{(2)}$ in differente networks. From the distribution and extreme value of $\Delta \lambda^{(2)}/\lambda^{(2)}$, we can know that the link's removal of nodes with smallest degree almost does not affect $\lambda^{(2)}$ of the network, $\Delta \lambda^{(2)} \ll \lambda^{(2)}$.}
\label{linkaccuracy}
\end{figure*}

Since $\vec{\xi}^{(i)tr}\vec{\xi}^{(i)}=1$, $\Delta \lambda^{(i)} \simeq -(\xi^{(i)}_k-\xi^{(i)}_m)^2$. Therefore, the perturbation of $\Delta \lambda^{(2)}$ mainly depends on the Fielder vector \cite{chung1997spectral} (eigenvector corresponding to the smallest non-negative eigenvalue $\lambda^{(2)}$). The Fiedler vector could be obtained by minimising the degree-adjusted Rayleigh quotient \cite{chung1997spectral} \cite{chung2002chip}. It is not difficult to find that $(\xi^{(i)}_k-\xi^{(i)}_m)^2 < 2$. Acutually, $(\xi^{(i)}_k-\xi^{(i)}_m)^2 \ll 1$, which means that $\Delta \lambda^{(2)} \ll \lambda^{(2)}$ (shown in Fig. (\ref{linkaccuracy})). Therefore, we can ignore the $\Delta \lambda^{(2)}$ and obtain $\lambda^{(2)}$ of the original network by estimating $\lambda^{(2)}$ of the new network according to
\begin{equation}\label{equ18}
    \lambda^{(2)} \simeq \min{\bigg((k_i-1) +  \sum_{j \ne i}\frac{A_{ij}^2}{(k_i-1)-k_{j}} \bigg|_{k_i=k_{\rm min} }\bigg)},
\end{equation}
where $k_{\rm min}$ is the minimum degree of the original network, $A_{ij}$ represents the connections of node $i$.

In a given network, degree of nodes is determined, so $\lambda^{(2)}$ is mainly affected by $\sum_{i \ne j}\frac{A_{ij}^2}{(k_i-1)-k_{j}}$. We assume that there are $z$ nodes with minimum degree $k_{\rm min}$ in the original network. Since these $z$ nodes have their own different connections, $\sum_{i \ne j}\frac{A_{ij}^2}{(k_i-1)-k_{j}}$ varies according to different nodes. So $\lambda^{(2)}$ is determined by the node with smallest $\sum_{i \ne j}\frac{A_{ij}^2}{(k_i-1)-k_{j}}$ of these $z$ nodes. (\ref{equ18}) shows that connections of nodes with smallest degree mainly affect $\lambda^{(2)}$, otherwise, nodes with large degree does not have much effects on $\lambda^{(2)}$. That is to say node with smallest degree connect to nodes with similar degree will decrease $\lambda^{(2)}$. This implys that there exists relationship between local assortativity and $\lambda^{(2)}$. Local assortativity is a property of a single node and indicates how similar a node is to its neighbours \cite{piraveenan2008local}. A lot of methods are proposed to calculate local assortativity. A simple method proposed in \cite{thedchanamoorthy2014node} to calculate local assortativity $\rho_i$ is used in this paper:
\begin{equation}\label{equ19}
    \rho_i = \frac{r+1}{N}-\bar{\theta_i},
\end{equation}
where $\bar{\theta_i} = \theta_i/\sum_i^N\theta_i$, $r$ is the assortativity of the network and $\theta_i$ is calculated by
\begin{equation}\label{equ20}
    \theta_i = \frac{1}{k_i}\sum_{i=1}^NA_{ij}\left|k_i-k_j \right|.
\end{equation}

\begin{figure}[htbp]
\centering
\resizebox*{9cm}{!}{\includegraphics{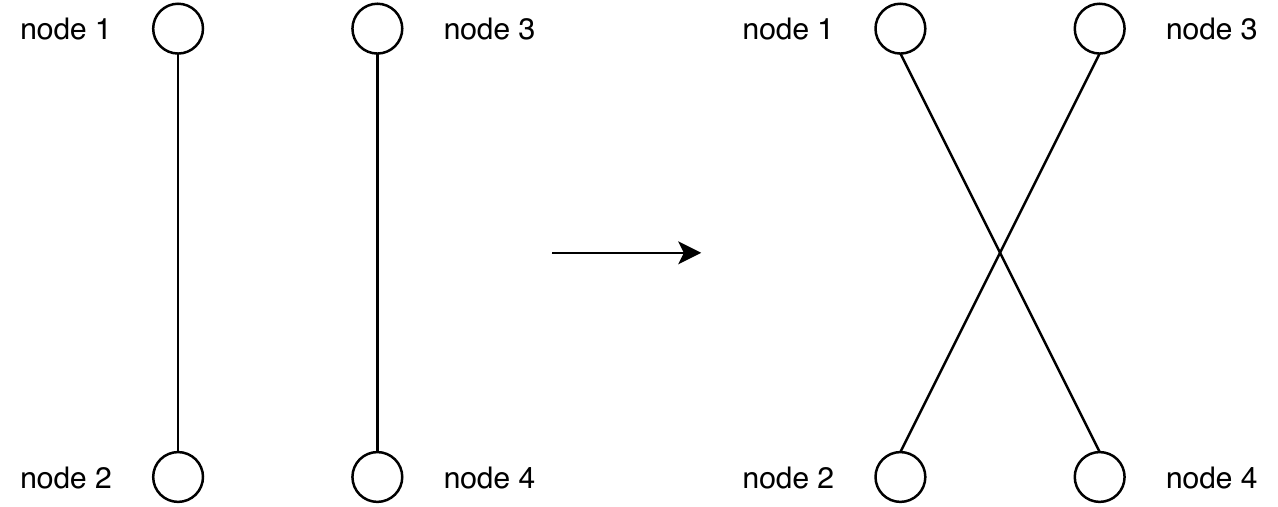}}
\caption{Rewire connections between nodes. This method include two steps: delete connections and add new connections. By this method, the neighbour of these four nodes have been changed but their degree. So, we can use this method to adjust the network structure while maintaining the degree of each node.}
\label{reconnection}
\end{figure}

\begin{figure*}[ht]
\centering  
\subfigure[estimation of $\lambda^{(N)}$]{ 
\begin{minipage}{0.4\textwidth}
\centering    
\includegraphics[scale=0.5]{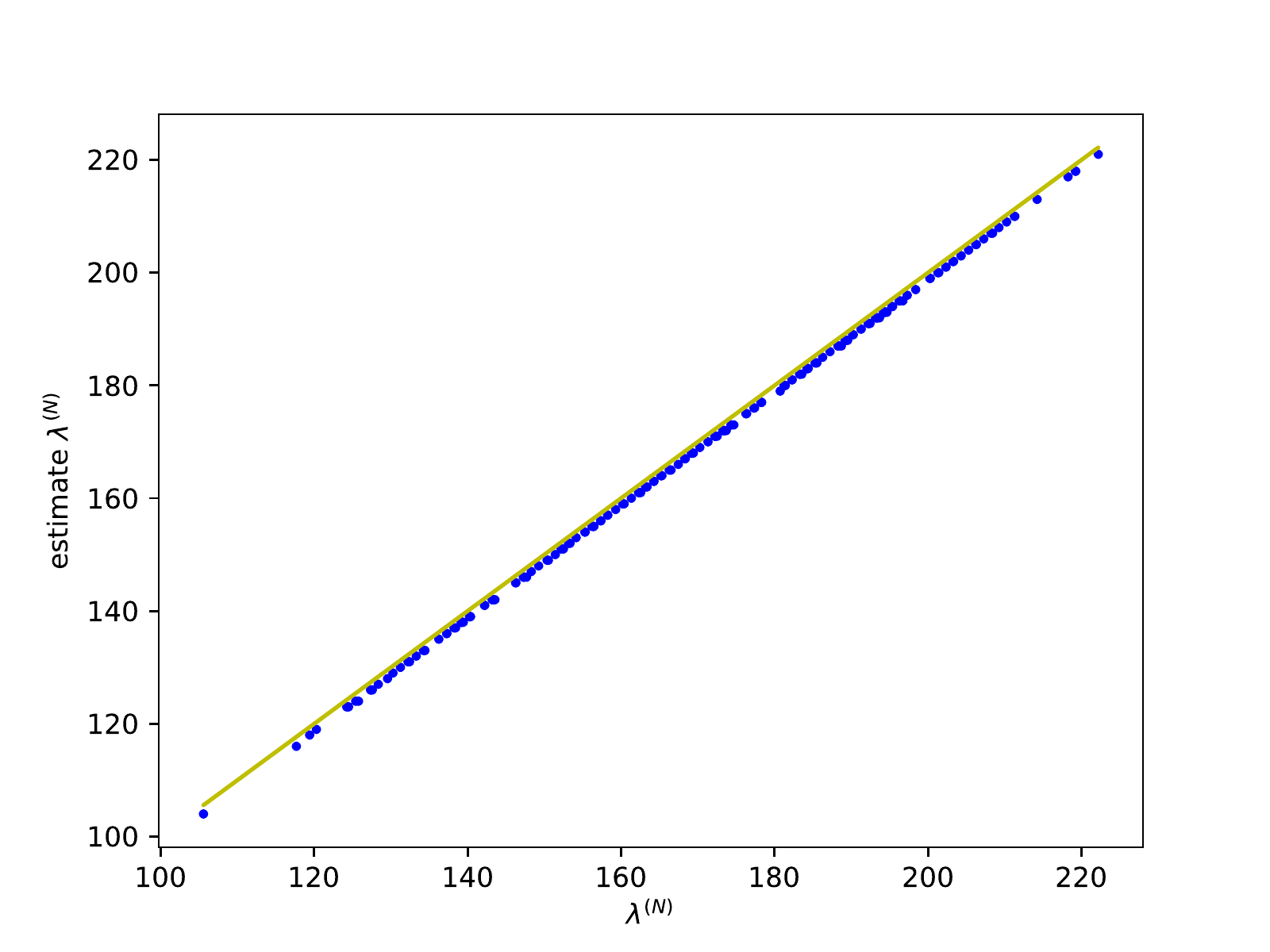}
\end{minipage}
}
\hspace{0.5in}
\subfigure[estimation of $\lambda^{(2)}$]{
\begin{minipage}{0.4\textwidth}
\centering   
\includegraphics[scale=0.5]{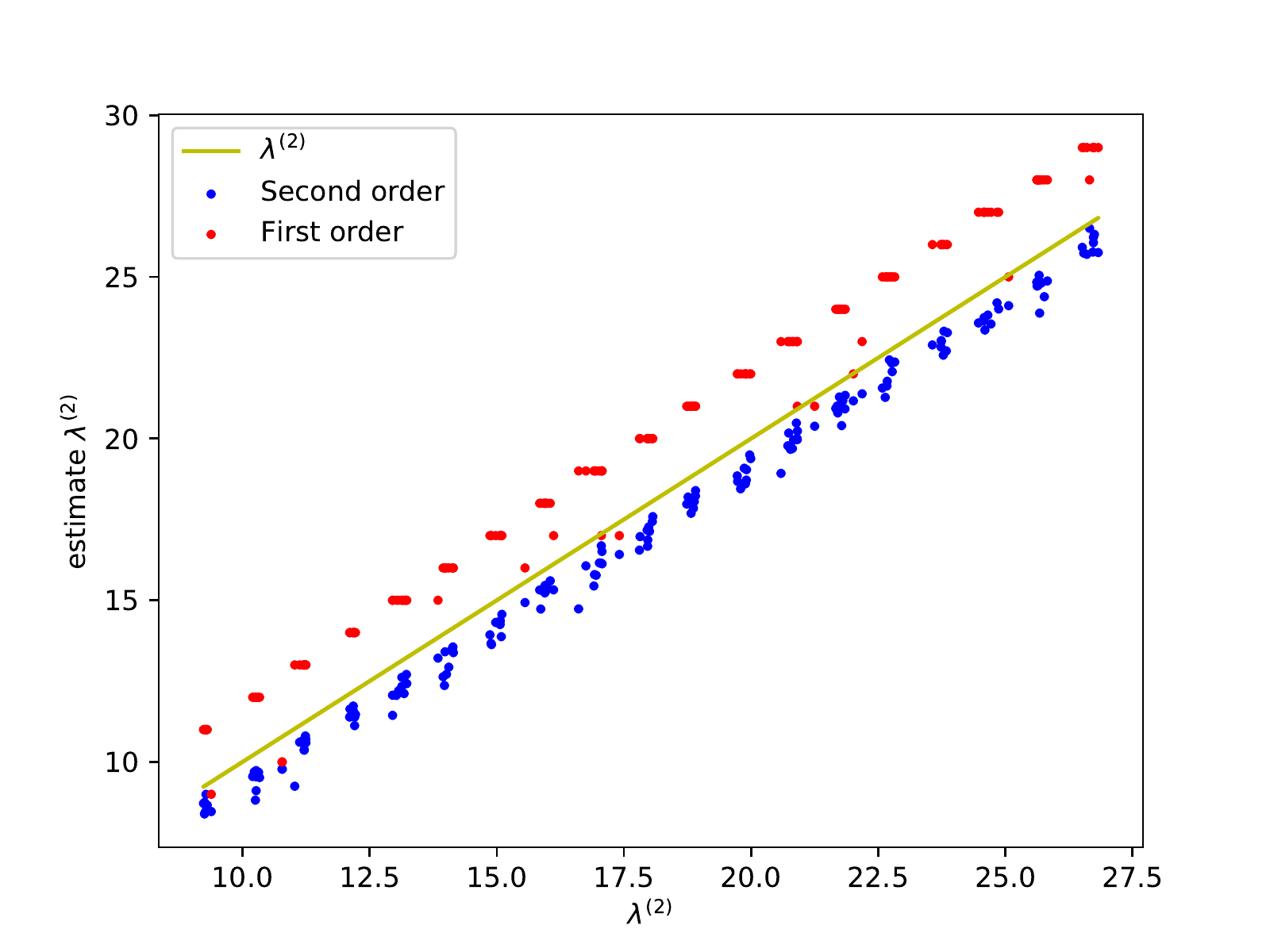}
\end{minipage}
}
\caption{(a) shows that the estimation of $\lambda^{(N)}$, $k_N+1$, is very close to the real value of $\lambda^{(N)}$. Therefore, $k_N+1$ is a good estimation of $\lambda^{(N)}$ is SF network. (b) shows the estimation of $\lambda^{(2)}$ different SF netwoks by purterbation theory method proposed in this paper. The red and blue scatter graph represents the first and second order estimation of $\lambda^{(2)}$. The yellow line represents the real values of $\lambda^{(2)}$. In most cases, the second order estimation is closer to the real value of $\lambda^{(2)}$ than the first order.}   
\label{estaccuracy}    
\end{figure*}

In (\ref{equ19}), we can see that $\rho_i$ is determined by $\frac{r+1}{N}$ and $\bar{\theta_i}$. Generally speaking, similar connection (connection between similar degree nodes) will increase $r$. The effect of similar connection on $\bar{\theta_i}$ need to be analyzed. In a large scale SF network with $N$ nodes, where $N$ is a large number, we maintain the degree sequence of the network and only rewire the connections of nodes shown in Fig. (\ref{reconnection}). It is assumed that node $1$ and node $4$ have small degree, while node $2$ and node $3$ have large degree. $k_1 \simeq k_4 \ll k_2 \simeq k_3$. So, connection between node $1$ and node $4$ as well as connection between node $2$ and node $3$ could be seen as similar connection. Since we analyse the local assortativity of the node with smallest dgree, we assume $k_1 = k_{\rm min}$. Let $\theta_{\rm sum} = \sum_i^N\theta_i$. The change of $\theta_{\rm sum}$ from left conections to right connections in Fig (\ref{reconnection}) is $\Delta \theta_{\rm sum}$ and the change of $\theta_1$ is $\Delta \theta_1$. In the left connections,  $\bar{\theta_1}=\theta_1/\theta_{\rm sum}$; in the right connections, $\bar{\theta_1'}=(\theta_1+\Delta \theta_1)/(\theta_{\rm sum}+\Delta \theta_{\rm sum})$. Then  
\begin{equation}
    \begin{split}
         \Delta \bar{\theta_1} &= -\bar{\theta_1}+ \bar{\theta_1'}
         = \frac{-\theta_1}{\theta_{\rm sum}} + \frac{\theta_1+\Delta\theta_1 }{(\theta_{\rm sum}+\Delta \theta_{\rm sum})} \\
         &= \frac{\theta_{\rm sum}\Delta \theta_1-\theta_1\Delta \theta_{\rm sum}}{(\theta_{\rm sum}+\Delta\theta_{\rm sum})\theta_{\rm sum}};
    \end{split}
\end{equation}

\begin{equation}
    \Delta \theta_1 = \left| \frac{k_4-k_1}{k_1}\right|-\left| \frac{k_2-k_1}{k_1}\right| \simeq -\left| \frac{k_2-k_1}{k_1}\right| < 0;
\end{equation}

\begin{equation}
    \begin{split}
        &\Delta \theta_{\rm sum} = (\left| \frac{k_4-k_1}{k_1}\right| + \left| \frac{k_4-k_1}{k_4}\right|+ \left| \frac{k_3-k_2}{k_3}\right| + \left| \frac{k_3-k_2}{k_2}\right|)\\
        &- (\left| \frac{k_2-k_1}{k_1}\right| + \left| \frac{k_2-k_1}{k_2}\right|+ \left| \frac{k_3-k_4}{k_3}\right| + \left| \frac{k_3-k_4}{k_4}\right|) \simeq 0.
    \end{split}
\end{equation}
Since $\Delta \theta_{\rm sum}>0$ and $\Delta \theta_1<0$, we can get $\Delta \bar{\theta_1}<0$. The simliar connection cause the increase of $\rho_i$ and the decrease of $\lambda^{(2)}$, which means a negative relation exists between local assortativity and $\lambda^{(2)}$.

\begin{figure}[htbp]
\centering
\resizebox*{8cm}{!}{\includegraphics{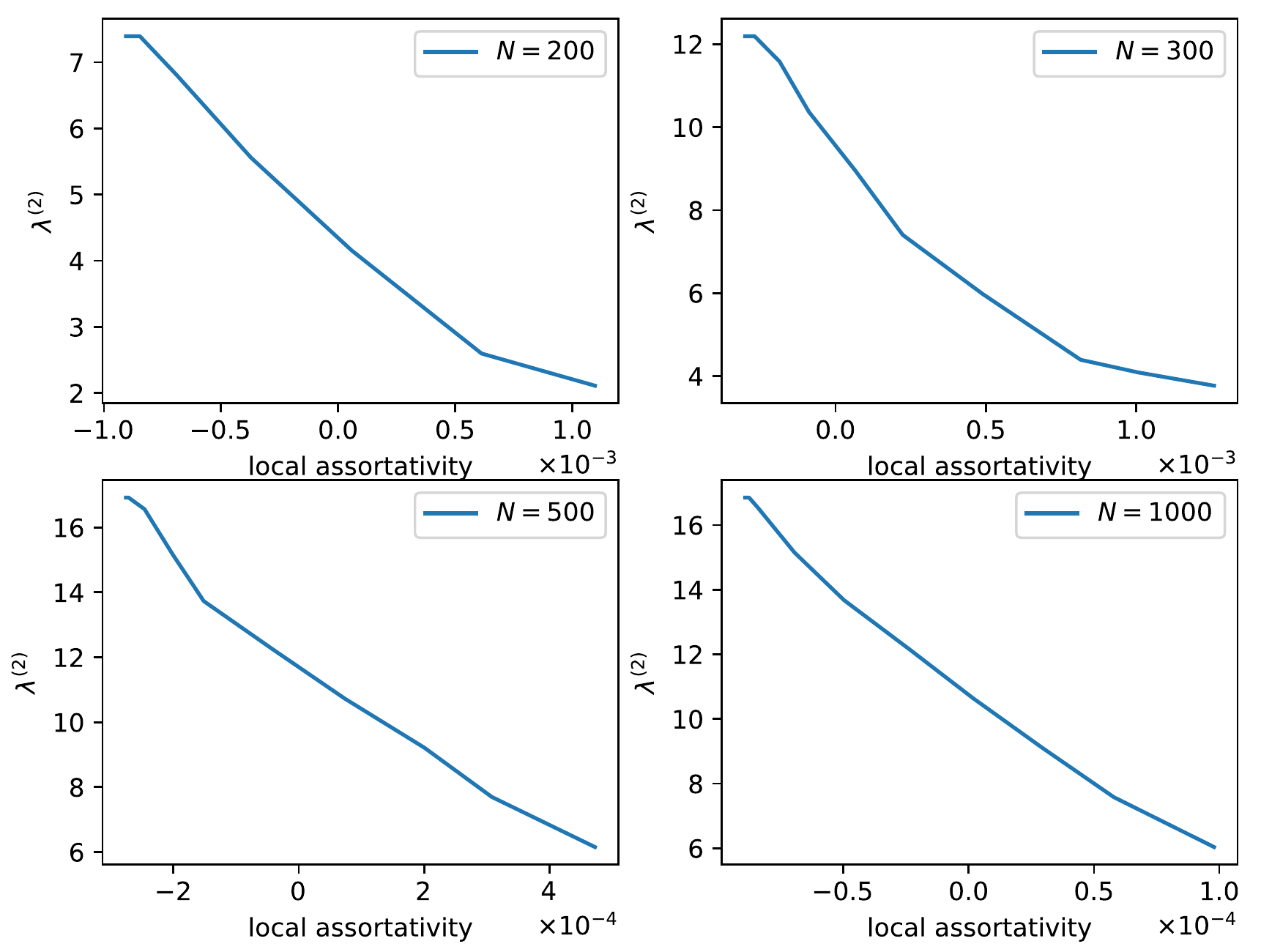}}
\caption{Relationship between local assortativity and $\lambda^{(2)}$ in BA networks with different size. All of these figures show that $\lambda^{(2)}$ decreases as the increase of the local assortativity.}
\label{assortativity}
\end{figure}

\section{Results and analysis} \label{sec:results}

\begin{figure*}[htbp]
\centering
\resizebox*{16cm}{!}{\includegraphics{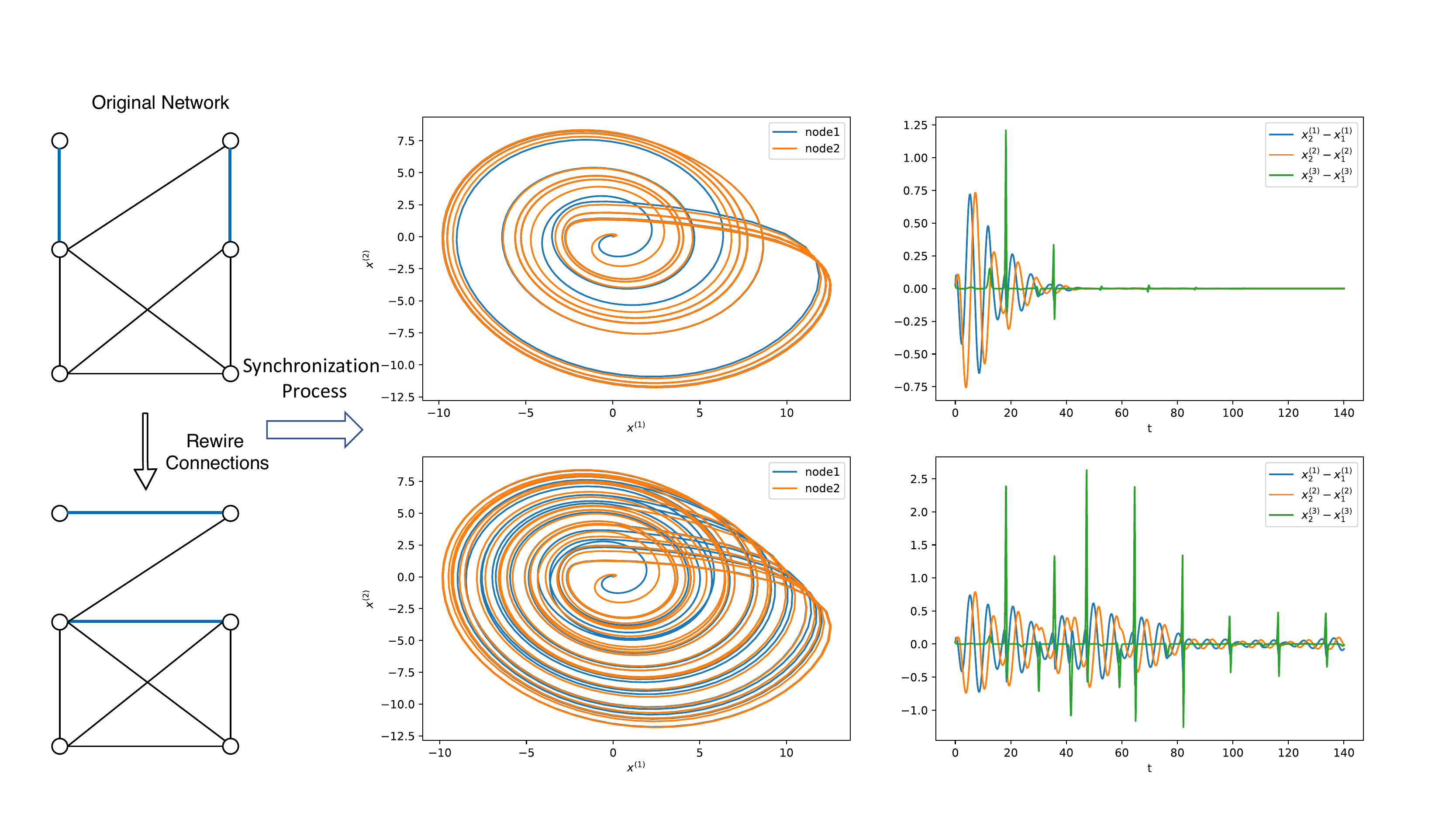}}
\caption{Control the synchronizability of SF network by rewiring connections of nodes with smallest degree. From the top three subfigures, we can see that the orignal network has a good synchronizability and the ndoes' trajectories eventually synchronize. In the three bottom subfigures, it shows that the network structure has been adjusted by rewiring connections. The node with the smallest degree connects to similar nodes. The trajectories of these node do not achieve synchronization at last.}
\label{desyn}
\end{figure*}

Here, we consider Rossler system \cite{rossler1976equation} for our nodes. The dynamic function of the $ith$ oscillator is 
\begin{equation}\label{equ21}
    \begin{split}
        &\dot{x_{i}}^{(1)}=-({x_{i}}^{(2)}+{x_{i}}^{(3)})\\
        &\dot{x_{i}}^{(2)}={x_{i}}^{(1)}+a{x_{i}}^{(2)}\\
        &\dot{x_{i}}^{(3)}=b+{x_{i}}^{(3)}({x_{i}}^{(1)}-d).
    \end{split}
\end{equation}

We assume that $a=0.2, b=0.2, d=6.0$, the coupling strength $c=0.03$. Also, nodes are coulped by $x^{(1)}$. The inner coupling function $H$ is
\begin{equation}\label{equ22}
    \begin{bmatrix}
   1 & 0 & 0 \\
   0 & 0 & 0 \\
   0 & 0 & 0
  \end{bmatrix}.
\end{equation}
According to master stability function, the stability region is: $S= (-5.44,-0.199)$.  

%
%\begin{figure}[htbp]
%\centering
%\subfigure[]{
%\resizebox*{9cm}{!}{\includegraphics{picture/estimation_accuracy_maxeigen.pdf}}}\hspace{5pt}
%\subfigure[]{
%\resizebox*{9cm}{!}{\includegraphics{picture/estimation_accuracy.pdf}}}
%\caption{(a) shows that the estimation of $\lambda^{(N)}$, $k_N+1$, is very close to the real value of $\lambda^{(N)}$. Therefore, $k_N+1$ is a good estimation of $\lambda^{(N)}$ is SF network. (b) shows the estimation of $\lambda^{(2)}$ different SF netwoks by purterbation theory method proposed in this paper. The red and blue scatter graph represents the first and second order estimation of $\lambda^{(2)}$. The yellow line represents the real values of $\lambda^{(2)}$. In most cases, the second order estimation is closer to the real value of $\lambda^{(2)}$ than the first order.} 
%\label{estaccuracy}
%\end{figure}
%

We consider different size of SF network generated by Barabási–Albert (BA) model \cite{barabasi1999emergence}. The estimation method proposed above will be used to estimate the extreme eigenvalues and the accuracy of the method will be verified by simulations. In Fig. (\ref{estaccuracy}), it shows that $k_{N}+1$ is very close to the real value $\lambda^{(N)}$ in different SF network. Therefore, $k_{N}+1$ is a good estimation of $\lambda^{(N)}$. Estimation of $\lambda^{(2)}$ by the proposed method is shown in Fig. (\ref{estaccuracy}). Comparing with the first-order estimation, the second-order estimation is closer to the real value of $\lambda^{(2)}$. That is say, in most situation, (\ref{equ18}) is a better estimation of $\lambda^{(2)}$ than $k_{\rm min}$ and  $\lambda^{(2)} \lesssim k_{\rm min}$.

To verify the relationship between local assortativity, the rewire connections method in Fig. (\ref{reconnection}) is applied in BA networks. Fig. (\ref{assortativity}) shows the relationship between local assortativity and $\lambda^{(2)}$ in different BA networks with $N=200, 300, 500, 1000$. $\lambda^{(2)}$ decreases with the increase of local assortativity. Since the rewire connections method does not change the degree sequence of the network, $\lambda^{(N)} \simeq k_{N}+1$ will not change according to the above analysis. Therefore, the ratio of $\frac{\lambda^{(N)}}{\lambda^{(2)}}$ will increase with the local assortativity, which indicates the decrease of synchronizability of the network.

Therefore, (\ref{equ18}) provides us a good strategy to control the synchronizability while maintaining the degree sequence of the network. If the node with smallest degree connect to nodes with similar degree, $\lambda^{(2)}$ will decrease and this action will weaken the synchronizability. On the other hand, if the node with smallest degree connect to nodes with large degree, $\lambda^{(2)}$ will increase and the synchronizability will be strengthen. Here, a BA network with Rossler dynamics described in (\ref{equ21}) (\ref{equ22}) will be used to verify the effectiveness of the strategy. The network of 100 nodes is grown by attaching new nodes with 10 edges that are preferent attached to existing nodes with high degree. In this network $G1$, $\lambda^{(N)}=45.6$ and $\lambda^{(2)}=7.6$. $-c\lambda^{(2)}=-0.228 \in S, -c\lambda^{(N)}=-1.368 \in S$. Then, for every eigenvalue of this network, $-c\lambda^{(i)} \in S$. According to the above analysis before, every nodes of this network will achieve synchronization (The three pictures at the bottom of Fig. (\ref{desyn}) show the synchronization process of nodes). Then, according to the above controlling strategy, we rewire connections of the node with similar nodes and get a new network $G2$. The extreme eigenvalues of $G2$, $\lambda^{(2)}=2.6$ and $\lambda^{(N)}=45.6$. Then, $-c\lambda^{(N)} \in S$ but $-c\lambda^{(2)} \notin S$. Therefore, nodes in $G2$ can not achieve global synchronization (the top three pictures in Fig. (\ref{desyn}) show the synchronization process of some nodes in $G2$). Therefore, rewiring connections of nodes with smallest degree is a valid strategy to enhance or weaken the synchronization of the whole network.

\section{Conclusions} \label{sec:conclusions}
In this paper, we have proposed an analytical method to estimate extreme eigenvalues of Scale-free(SF) network. To avoid the degeneration of eigenvectors when estimating the smallest non-zero eigenvalue of SF network, a link removal method has been used. Then the non-degenerate perturbation theory can be used to do the estimation and only requires the local information of smallest nodes in the network. The non-degenerate perturbation theory can give us an analytic equation of estimation, which indicates that there exists a negative relationship between the smallest non-zero eigenvalue $\lambda^{(2)}$ and the local assortativity of the smallest node. Also, the equation informs us how to control the synchronizability of the network by rewiring the connections of smallest nodes. By simulation of Rossler system in different SF networks, the method has been verified. Therefore, this paper helps us understand the relationship between the connections of smallest nodes and the global synchronizability of the network. And also it can inform us how to control the synchronizability of the network by rewiring connections of nodes while maintaining the degree sequence of the network. In the future, we expect to explore what causes the difference between estimation value and the true value of extreme eigenvalue. The difference may reveal potential network topology that affects global synchronizability.
\section*{Author Contributions} \label{sec:contribution}
Mengbang Zou performed simulation. Mengbang Zou and Weisi Guo together wrote the paper, analysed and discussed the results.
\bibliographystyle{apsrev4-1}
\bibliography{Ref}
%\begin{thebibliography}{4}
%\bibitem{Griffiths}
%D. J. Griffiths,
%\textit{Introduction to Electrodynamics}
%(Cambridge University Press, Cambridge, 2017).

%\bibitem{Fleming}
%A. Bobrinha,
%Revista Brasileira de Lorem Ipsum \textbf{23},
%179 (2002).

%\bibitem{Feynman}
%R. P. Feynman, R. B. Leighton and M. Sands,
%\textit{Lições de Física de Feynman}
%(Editora Bookman, Porto Alegre, 2008).

%\bibitem{Jackson-CE}
%J. D. Jackson,
%\textit{Classical Electrodynamics}
%(John Wiley \& Sons, Danvers, 1999).
%\end{thebibliography}

%\appendix*
%\input{sections/appendix1.tex}

\end{document}